# A liquid nitrogen-cooled Ca$^+$ optical clock with systematic uncertainty of 3×10$^{-18}$

Yao Huang[1,2], Baolin Zhang[1,2,3], Mengyan Zeng[1,2,4], Yanmei Hao[1,2,3], Huaqing Zhang[1,2,3], Hua Guan[1,2], Zheng Chen[1,2,3], Miao Wang[1,2,3], & Kelin Gao[1,2]

[1]*State Key Laboratory of Magnetic Resonance and Atomic and Molecular Physics, Innovation Academy for Precision Measurement Science and Technology, Chinese Academy of Sciences, Wuhan 430071, China*
[2]*Key Laboratory of Atomic Frequency Standards, Innovation Academy for Precision Measurement Science and Technology, Chinese Academy of Sciences, Wuhan 430071, China*
[3]*University of Chinese Academy of Sciences, Beijing 100049, China*
[4]*Huazhong University of Science and Technology, Wuhan 430074, China*

Here we present a liquid nitrogen-cooled Ca$^+$ optical clock with an overall systematic uncertainty of 3×10$^{-18}$. In contrast with the room-temperature Ca$^+$ optical clock that we have reported previously, the temperature of the blackbody radiation (BBR) shield in vacuum has been reduced to 82(5) K using liquid nitrogen. An ion trap with a lower heating rate and improved cooling lasers were also introduced. This allows cooling the ion temperature to the Doppler cooling limit during the clock operation, and the systematic uncertainty due to the ion's secular (thermal) motion is reduced to <1×10$^{-18}$. The uncertainty due to the probe laser light shift and the servo error are also reduced to <1×10$^{-19}$ and 4×10$^{-19}$ with the hyper-Ramsey method and the higher-order servo algorithm, respectively. By comparing the output frequency of the cryogenic clock to that of a room-temperature clock, the differential BBR shift between the two was measured with a fractional statistical uncertainty of 7×10$^{-18}$. The differential BBR shift was used to calculate the static differential polarizability, and it was found in excellent agreement with our previous measurement with a different method. This work suggests that the BBR shift of optical clocks can be well suppressed in a liquid nitrogen environment. This is advantageous because conventional liquid-helium cryogenic systems for optical clocks are more expensive and complicated. Moreover, the proposed system can be used to suppress the BBR shift significantly in other types of optical clocks such as Yb$^+$, Sr$^+$, Yb, Sr, etc.

Optical clocks have higher accuracy and stability than microwave clocks and may potentially be used to update the SI definition of a second [1]. In addition to optimizing the precision of time and frequency standards, optical clocks find applications in various fields such as quantum physics [2, 3], tests of general and special relativity [4], geoid measurements [5], the physics beyond the standard model [6, 7], etc. Optical clocks with better fractional frequency uncertainty and stability, at the 10$^{-18}$ or even better, would help improving the testing/measuring sensitivity, thus would increase the chance for searching new physics.

Since the introduction of optical clocks, significant advances have been realized in decreasing the uncertainty and improving the stability [8]. Current high-performance optical clocks can be categorized into two types: (a) optical lattice clocks (OLCs) referenced to atoms trapped in an optical lattice [9-11], and (b) clocks referenced to a single ion trapped in a radio frequency (RF) trap [12-15]. The optical clocks referenced to Al$^+$ [12], Sr [9, 10], Yb$^+$ [14], and Yb [11] have reach systematic uncertainties at the 10$^{-18}$ level. In particular, the Al$^+$ clock made by NIST has reached a systematic uncertainty of <1×10$^{-18}$ [12]. However, most of clocks exhibit fractional BBR shifts >1×10$^{-15}$ at room temperature [9, 11, 14, 15], excluding systems using certain optical standards such as Al$^+$ that are not sensitive to BBR. In many state-of-the-art single-ion optical clocks and OLCs, the total systematic uncertainty is often dominated by BBR shift uncertainties [9,11,15]. Therefore, reducing the BBR shift is crucial in realizing high-precision atomic clocks.

For an atomic clock, the BBR shift would be [16]

$$\Delta \nu_{BBR} = -\frac{\Delta\alpha(0)}{2h} \langle E^2 \rangle_T [1 + \eta(T)] \qquad (1)$$

where $\Delta\alpha(0)$ is the differential scalar polarizability for the clock transition, $h$ is the Plank's constant, $\langle E^2 \rangle_T$=[8.319430(15) V/cm]$^2$(T/300 K)$^4$ is the mean-squared electric field in a BBR environment at the temperature $T$ [17], $\eta(T)$ gives a small dynamic correction to the BBR shift. The evaluation of $\Delta\nu_{BBR}$ needs both the knowledge of $\Delta\alpha(0)$, $\eta(T)$ and the precise evaluation of the BBR environmental temperature $T$.

A possible approach to address these issues involves choosing optical standards that are not sensitive to BBR shifts. In particular, the BBR shift sensitivity for the clock transition in Al$^+$ is over hundreds of times smaller than that of other optical clocks like Sr, Yb, Sr$^+$, Yb$^+$, or Ca$^+$ [9, 11, 12, 14, 15, 18], and it can be reduced to a low level even at room temperature. Previous studies also have been investigated using Hg [19], In$^+$ [20], and Lu$^+$ [21] clocks with low sensitivity to BBR. Highly-charged-ion-based optical clocks with negligible sensitivity to BBR have also been studied [22-24]. However, for many of the systems described above, they often require using deep-ultraviolet light sources; for some of the clocks, they can only be achieved with quantum logic spectroscopy [12].

Another approach to realize accurate reference systems for optical clocks is to accurately measure the sensitivity parameters $\Delta\alpha(0)$ and $\eta(T)$, in the meantime, precisely measuring the ambient temperature $T$ and thus giving a precisely evaluated fractional BBR shift. However, evaluating the uncertainties in the fractional BBR shift in the 10$^{-18}$ level at room temperature requires very high temperature accuracy, even when $\Delta\alpha(0)$ and $\eta(T)$ have already been known with a very high accuracy. Evaluating the uncertainties in the fractional BBR shift in the 10$^{-18}$ level at room temperature is challenging and requires either the dynamic temperature calibration [9] or BBR shields with accurate systematic error analysis [25]. In particular, the RF fields used to trap ions in single-ion-based optical clocks would heat the ion traps, which results in an uneven thermal environment, making the BBR shift estimation even more difficult.

As stated in Eq. (1), the BBR shift is proportional to the fourth power of the temperature ($T^4$). Thus, decreasing the temperature of the BBR field can reduce the BBR shift and the corresponding temperature dependence ($\frac{\partial \nu_{BBR}}{\partial T} \propto T^3$). For example, the BBR shift for $^{40}$Ca$^+$ at room temperature is ~345 mHz [26], an evaluated temperature uncertainty of $\Delta T \approx 0.1$ K yields a fractional uncertainty at $1 \times 10^{-18}$, while the BBR shift would be reduced to only 1.6 mHz at the liquid nitrogen temperature of ~77 K. The fractional uncertainty can be reduced to $1 \times 10^{-18}$, as long as the temperature is within an uncertainty of 4 K. The cryogenic environment has been used in the Hg$^+$ optical clock, the Sr OLC, and the Cs fountain clock, which greatly suppresses the BBR shift and its uncertainty [10, 13, 27].

In this work, we report a liquid nitrogen-cooled Ca$^+$ optical clock with an overall systematic uncertainty of $3.0 \times 10^{-18}$, which is approximately an order of magnitude lower than that for our previously reported Ca$^+$ optical clocks at room temperature [18, 26]. The differential BBR shift between the two systems was determined by comparing the corresponding clock frequencies. The differential BBR shift was further used to derive the static differential polarizability, which was consistent with our previously reported result [18]. This provides a test for the measured $\Delta\alpha(0)$ with a different method.

Ca$^+$ is one of the candidates that have high-Q optical transitions, it's natural linewidth is ~0.14 Hz, suitable for building a high-performance optical clock. Its level scheme is relatively simple, only low-cost diode lasers would be needed. As a result, portable and robust clock can be made [26], these features would broaden the application for the optical clocks. In our previous work, both the laboratory Ca$^+$ optical clocks [18] and the portable Ca$^+$ optical clocks [26] have been made, with systematic uncertainty at the $10^{-17}$ level. The excess micromotion induced 2$^{nd}$-order Doppler shift and Stark shift, the Stark shift due to the secular motion, and the 2$^{nd}$-order Doppler shift due to the secular motion (micromotion induced) have been canceled by choosing the magic RF trapping frequency [15, 18]. The quadrupole shift, the 1$^{st}$-order Zeeman shift, and the tensor Stark shifts due to the ion motion and the lasers have been canceled by averaging the 3 pairs of Zeeman transitions [28-30]. For the previous made optical clocks, the total systematic uncertainty is limited by the BBR shift, the 2$^{nd}$-order Doppler shift due to the secular motion, and the ac Stark shift (light shift), etc.

In this article, we introduce a recently built, liquid nitrogen-cooled Ca$^+$ optical clock, for which the liquid nitrogen environment greatly reduced the BBR shift and its uncertainties. A new ion trap is introduced for lower heating rates. In addition, cooling lasers under an optimized working condition were adopted for cooling the ion trap close to the Doppler cooling limit. This reduces the 2$^{nd}$-order Doppler shift due to the ion's secular motion. The hyper-Ramsey interrogation scheme [14, 31] was adopted to reduce the uncertainties in the frequency shift induced by the probe beam. Optimized servos were adopted to reduce servo-introduced uncertainties. Moreover, the 1$^{st}$-order Doppler shift was eliminated by using two probe beams in opposite directions for the detection [12].

A BBR shield and a special designed vacuum system were constructed for creating a liquid nitrogen environment to reduce the BBR shift and its uncertainty (Fig.1). The BBR shield was attached to the bottom of the liquid nitrogen container. The materials and the structure for designing the vacuum system is very important: we have tried a stainless-steel container, but > 200 μm of vertical displacement for the ion trap was observed during the liquid nitrogen consumption, making it impossible for running the clock experiments. Both the shield and the liquid nitrogen container were constructed from oxygen-free copper with good thermal conductivity. The liquid nitrogen container was approximately 0.8 m in height and 5 mm in thickness. The good thermal conductivity of the liquid nitrogen container ensured vertical temperature uniformity during liquid nitrogen consumption, thus avoiding changes in the height of the ion trap. In addition, three polytetrafluoroethylene fixing rods were attached to the bottom of the liquid nitrogen container to prevent horizontal displacement. The displacement of the ion trap would be < 10 μm for ~ 1.5 h after each round of filling of liquid nitrogen, ready for the optical clock experiments. The shield was equipped laterally with six windows of diameter 10 mm for light transmission, and a window of diameter 40 mm was installed at the bottom for both light transmission and ion fluorescence collection. The windows were made of antireflection coated BK7 glass. The transmittance of the BK7 glass was ~0 for wavelengths above 3 μm [25], thereby eliminating the effect of window transmittance on the BBR at room temperature. Three platinum resistance temperature probes were installed in the shield chamber, one at the top, one at the bottom, and one near the ion trap for the temperature measurement. There are eight holes of diameter 5 mm on the shield for keeping the vacuum environment for the ion trap.

The vacuum chamber was made of stainless steel, and three pairs of coils were installed outside the chamber for adjusting the strength and direction of the magnetic field. To achieve a more stable magnetic field environment for $Ca^+$, four layers of magnetic shielding were added. However, the magnetic field fluctuation would still broaden the observed clock transition, thus degrading the clock locking performance. Optimization has been made through fine adjustment of the magnetic field directions. Since the magnetic field fluctuation in the vertical direction is much stronger than that in the horizontal directions, the magnetic field direction is set to be horizontal, and the overall magnetic field amplitude is less sensitive to the vertical environmental magnetic field fluctuations.

The temperature inside the BBR-shielding varied between 77 K and 80 K during the consumption of liquid nitrogen. Considering the solid angle ratio of the small holes and the influence of the heat conduction of several thick wires of thickness ~1 mm, the BBR shift was estimated to be 3.0(1.1) mHz (see more details in Methods). Accordingly, the temperature-associated fractional uncertainty in the BBR shift was determined to be $2.7 \times 10^{-18}$.

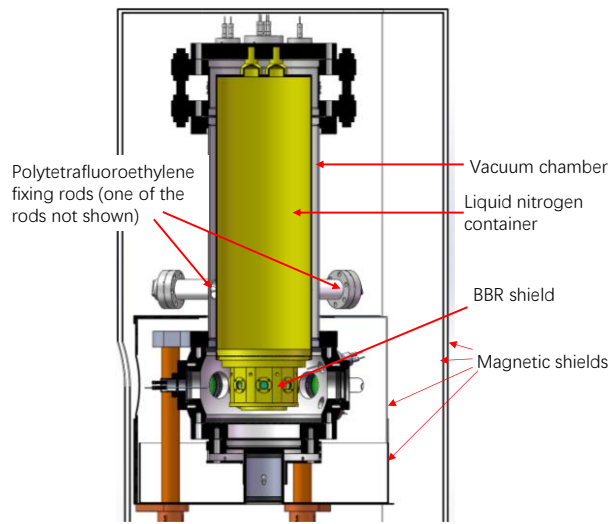

Fig. 1. The schematic diagram of the vacuum chamber used in the liquid nitrogen clock and the liquid nitrogen set up.

A diamond wafer based linear ion trap was used in the clock, which is similar to the NIST ion trap [12], but with larger size and slightly different structure. The highly symmetry and high precision laser machining help reducing the heating rates of the ion trap. The relatively open structure was also well suited for fluorescence detection at large solid angles, three-dimensional detection, and compensation for the micromotion of ions. The ion-trap heating rate was experimentally determined to be less than 1 mK/s, and less than 5 quanta/s for the frequencies of the radial secular motion. This is two orders of magnitude smaller than that of our previously reported ion trap [18,26]. In the experiment, both the 397-nm cooling beam and the 866-nm repumping beam were stabilized to an ultra-low expansion reference cavity, and the laser cooling of the ion was optimized by adjusting the frequencies of the acousto-optic modulators (AOMs). Multiple three-dimensional temperature measurements confirmed that the temperature of the Doppler-cooled ion was close to the Doppler cooling limit, and the temperature uncertainty was more than an order of magnitude lower than that in our previous room-temperature optical clock [18,26]. Considering the trap heating rates, the temperature was estimated to be 0.65-1.17 mK during the period for clock transition detection in the optical clock experiment, ie, $T_{ion}$= 0.91(26) mK. Accordingly, the 2$^{nd}$-order Doppler shift due to the secular motion was evaluated as -3.1(9)×10$^{-18}$.

Similar to the previous optical clock setup, we eliminated the electric quadrupole shift, the 1$^{st}$-order Zeeman shift, and the ion motion- and laser-induced tensor Stark shift by averaging the frequencies of 3 pairs of Zeeman transitions with different $^2D_{5/2}$ state sublevels [28-30]. However, the slow drift in the electric quadrupole shift over time might still lead to a residual uncertainty. A specific magnetic field direction is chosen to achieve a magic angle between the magnetic field and electric field gradients for reducing the drift rate of the electric quadrupole shift. This further reduced the residual uncertainty, leading to a reduction in the electric quadrupole shift by ~2 orders of magnitude. However, the three pairs of Zeeman sublevels had different Rabi frequencies in this specific magnetic field direction for the same probe-beam intensity. The stability of the optical clock was optimized by varying the power of the probe beam for different Zeeman sublevels. Accordingly, the ac Stark shift caused by the probe beam was also different for different Zeeman sublevels. Thus, the electric quadrupole shift could not be eliminated by simply averaging the frequencies of the three Zeeman pairs. Furthermore, the collimation of the probe beam would change over time, which introduces an additional systematic uncertainty. Therefore, it is important to suppress the ac Stark shift caused by the probe beam, which was greatly suppressed in the present experiment by adopting the hyper-Ramsey interrogation scheme [14,31]. The hyper-Ramsey interrogation scheme basically eliminates the optical frequency shift, and solves the above problems fundamentally. The fractional uncertainty of the frequency shift was estimated to be < 10$^{-19}$. For laser beams other than the probe beams, AOMs with mechanical shutters were used to eliminate the light shift [30].

When locking the frequency of the probe laser to the resonance of the clock transition, frequency drifts of the probe laser would cause a servo error, leading to a systematic frequency shift. The systematic frequency shift can be evaluated by statistically analyzing the quantum jump imbalances as an indicator of the servo shift and uncertainty [30]. In our experiment, a higher-order servo loop was used to dynamically lock quantum jump imbalances to 0. Moreover, the drift rate of the probe beam frequency was measured to be less than 63 mHz/s, and accordingly, the upper limit of servo-induced frequency shift was simulated to be 0.16 mHz, or 4×10$^{-19}$.

For our cryogenic clock, it is important for evaluating the 1$^{st}$-order Doppler shift since the ion trap still has a possible displacement during the clock running: the displacement might be very small and have a slow variation in speed, difficult to be observed. Besides, the photoelectric effect from laser beams may lead to changes in the stray electric field, which results in ion displacement in the ion trap. This in turn leads to a 1$^{st}$-order Doppler shift of up to order of 10$^{-17}$ when not suppressed [12]. In our experiment, two laser beams in opposite directions were used for independent and interleaved probing the ion, by averaging the two independent frequency measurements, the 1$^{st}$-order Doppler shift can be eliminated. The clock transition frequencies observed in our experiment with two counter-propagated probe beams differed from each other within 1×10$^{-17}$. The uncertainty in the 1st-order Doppler shift was estimated to be 3×10⁻¹⁹ with similar method to Ref [12].

There are a few other contributions for the systematic uncertainty, while they contribute < 1×10⁻¹⁹ to the total uncertainty. Table I summarizes the systematic uncertainty budget for the liquid nitrogen temperature clock. The total systematic uncertainty is 3.0×10⁻¹⁸, limited by the BBR temperature evaluation precision.

Table I. Systematic uncertainty budget for the liquid nitrogen temperature clock. Here only the effects with systematic shifts or uncertainty >1×10⁻¹⁹ is shown.

| Contribution | Fractional systematic shift ($10^{-18}$) | Fractional systematic uncertainty ($10^{-18}$) |
|---|---|---|
| BBR field evaluation (temperature) | 7.3 | 2.7 |
| BBR coefficient ($\Delta\alpha_0$) | 0 | 0.3 |
| Excess micromotion | 0 | 0.2 |
| 2nd-order Doppler (thermal) | -3.1 | 0.9 |
| Residual quadrupole | 0 | 0.4 |
| Servo | 0 | 0.4 |
| 1st-order Doppler | 0 | 0.3 |
| Total | 4.2 | 3.0 |

Since the determination of the blackbody environment is very important to the uncertainty budget of an optical clock, also for the verification of the reliability of our previous results [18], we further compared the frequencies of the cryogenic and the room-temperature optical clocks. Based on the frequency difference, the BBR shift could be determined directly, which could be used to further derive the sensitivity parameters $\Delta\alpha(0)$. The sensitivity parameters in our study were compared with those previously result for cross-verification. It was difficult to further improve the accuracy in our previous studies by using the magic trapping frequency measurements [18]. However, such a comparison is necessary, as it provides an independent method to verify the evaluation results.

The ion trap used in the room-temperature optical clock is similar to that in the NIST optical clock [12]. The diamond material and silver-plated oxygen-free copper support employed has high thermal conductivity, thereby facilitating heat conduction after RF heating of the ion trap. Furthermore, the ion-trap electrodes are coated with a thicker layer of gold to reduce the heating effect due to the RF field. In addition, the magic trapping frequency (~2π×24.8 MHz) allows the optical clock to work at low RF power (< 2 W). This further reduces the RF heating to the ion trap. However, fluctuations in room temperature cause the vacuum chamber temperature fluctuating by ±1.5 K, leading to a total systematic uncertainty of 1.7×10⁻¹⁷. The total systematic uncertainty is dominated by room temperature variations.

In the clock comparison experiment, a probe laser was used to synchronously probe both clocks [30]. However, both clocks were mutually independent in terms of the cooling, repumping, and quenching laser beams. Liquid nitrogen was completely consumed approximately 36 h after the liquid nitrogen container was filled, thus requiring the container to be filled once every day. This caused an interruption to the experiment for more than an hour every day. In addition to the systematic and statistical uncertainties, it is necessary to consider the gravitational redshift in the clock comparison experiment [5]. The two clocks were separated by about 8 m in the same laboratory with an ion-trap height difference of 15(1) cm as measured by a spirit level, equivalent to a frequency shift of 7.0(5) mHz or 1.7(1) ×10$^{-17}$. Fig. 2 shows the frequency comparison results, each data point shows the average of a measurement last for ~18 000 s, the error bar shows the calculated stability for the measurement. The systematic shifts other than the BBR shift, including the gravitational shift have been corrected in the figure. Taken our previous measured Δα(0) and calculated η(T) [18], for our Ca$^+$ clocks with cryogenic temperature and room temperature, the BBR shift difference would be calculated as -342(8) mHz, considering the gravitational shift of 7.0 (5) mHz and other systematic shifts, the frequency difference is expected to be -335(8) mHz. Averaged from data shown in Fig.2, considering both the systematic and the statistical shifts, the frequency difference is measured as -336(9) mHz, in excellent agreement to the above calculation. The differential scalar polarizability Δα(0) derived from the clock comparison would be -7.29(19)×10$^{-40}$ J m$^2$ V$^{-2}$, here the theoretical calculated dynamic correction η(T) is taken. The measurement lasted for 219 h, with a statistical uncertainty of 7×10$^{-18}$, which is within the systematic uncertainty of the room-temperature clock. Our previous measured Δα(0) by measuring the magic RF trapping frequency is -7.2677(21)×10$^{-40}$ J m$^2$ V$^{-2}$ [18], the two measurements are in excellent agreement. The experiment verified our previous measurement, also proved the reliability of the estimated BBR shift.

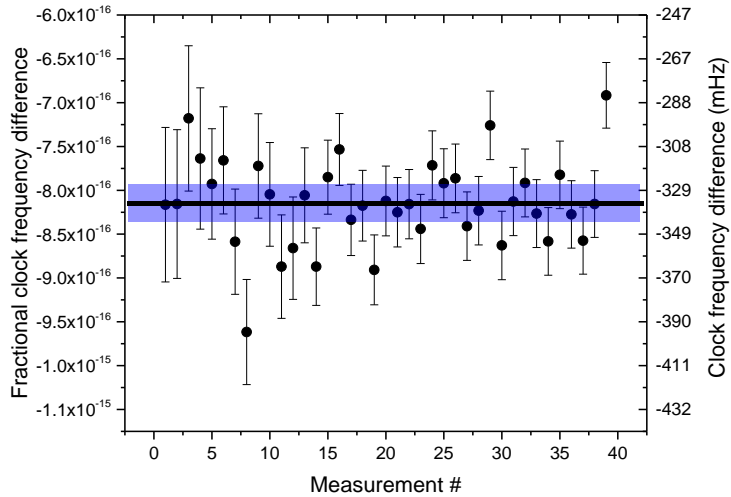

Fig.2. The measured fractional frequency difference between the cryogenic clock and the room temperature clock. The vertical axis represents the measured fractional frequency difference (left axis) and the frequency difference (right axis), the black solid line and the blue color band show the mean and the 1σ uncertainty of the measurement respectively, here both the statistical (7×10$^{-18}$) and the systematic uncertainties (2.1×10$^{-17}$) are included.

In summary, we have developed a liquid nitrogen-cooled Ca$^+$ clock with an uncertainty of the order of 10$^{-18}$. The differential BBR shift between the cryogenic clock (~82 K) and the room-temperature clock (~293 K) was directly determined through a comparative experiment, and the result was used to further derive the static differential polarizability Δα(0) of the Ca$^+$ clock transition.

The derived $\Delta\alpha(0)$ is in agreement to our previous measured value by measuring the magic RF trapping frequency. This work suggests that the BBR shift of optical clocks can be well suppressed in a liquid nitrogen environment. This is advantageous because conventional liquid-helium cryogenic systems for optical clocks are more expensive and complicated. Moreover, the proposed system can be used to suppress the BBR shift significantly in other types of optical clocks such as $Yb^+$, $Sr^+$, Yb, Sr, etc.

In the future, further simulation and analysis will be performed on the liquid-nitrogen cryogenic shield to further reduce the temperature evaluation uncertainty [32]. Once the total uncertainty is dominated by uncertainties in the 2$^{nd}$-order Doppler shift, it is necessary to further decrease the ion-trap temperature below the Doppler cooling limit with 3-dimentional sideband cooling [33] or electromagnetically induced transparency cooling [34]. The total uncertainty can be further reduced to within $1\times10^{-18}$. By introducing a probe laser with stability in the order of $1\times10^{-16}$ at 1-200 s, the probe time can be prolonged to ~ 1 s, it is possible to further improve the stability of $Ca^+$ optical clock, close to the single ion quantum projection noise limit [35]. The stability would be at the $10^{-18}$ level with an averaging time of a day. It is also possible to take multiple ions as a reference to further improve the clock stability. Moreover, we are planning to develop a new cryogenic system to achieve a longer continuous operation period. Developing a cryogenic system that can continuously work without interruption is also in consideration. Afterwards, clock comparison with other optical clocks would be made for testing the uncertainty evaluation and the measurement of the clock frequency ratios [36].

**Method**
**More details about the systematic uncertainty evaluation for the liquid nitrogen clock**
**Ion trap.** The ion trap we used in the experiment is shown in Fig. 3. The ion trap is similar to the one used in NIST $Al^+$ clocks [12], while with a modified design, the trap size is a bit larger. The ion trap is made of laser machined diamond wafer, while part of the diamond surface is gold plated to be used as trap electrodes, with a coating thickness of ~5 μm. The thickness of the diamond wafer is ~ 400 μm, while the distance between the RF electrodes and the trapped ion is ~ 400 μm, the two pairs of RF electrodes are designed fully symmetrical, for achieving smaller differential- RF -phase-induced micromotion and smaller trap RF heating rates. For NIST's design, two compensation electrodes are on the diamond wafer for the compensation of the micromotion. For making the RF electrodes fully symmetrical, in our design, only the horizontal compensation electrode is kept on the diamond wafer, while the vertical compensation is made by introducing compensation wires. For convenience of vertical laser detection through the ion trap, two 0.2-mm-thick copper wires are added for vertical compensation.

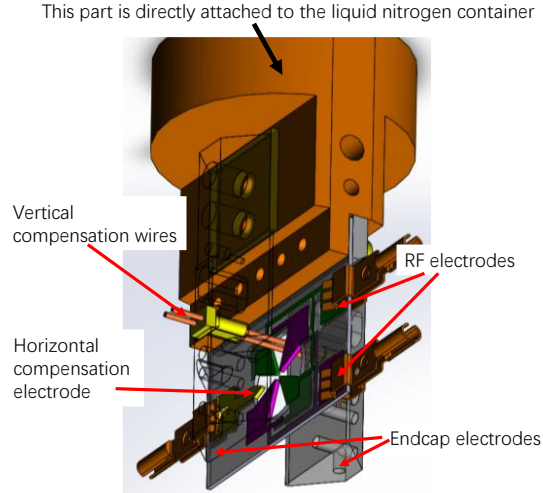

Fig. 3 The schematic diagram of the ion trap used in the liquid nitrogen clock.

**BBR shift.** Temperature sensors are installed on the trap wafer holder close to the trap, trap wafer holder ~ 10 cm away from the trap, and the BBR shield, however there is no sensors on the compensation wires and the endcaps, thin wires are connecting the endcaps to the feedthroughs on the vacuum chamber, whose temperature is ~293 K. Simulations are made for evaluating the temperature for the endcaps, showing a result of 77-87 K, results vary with different given emissivity and thermal conductivity. According to the temperature reading near the trap, between 77 and 80 K; for simplicity, we take 77 and 87 K as the temperature limits for the whole trap electrodes, the windows, and the BBR shield. There are eight small holes, would let in the blackbody radiation at room temperature, giving rise to the BBR shift; this effect is simulated, and the evaluated shift rise varies between 0.3 and 1.5 mHz with different assuming emissivity, we take these shift limits for the evaluation. Table II gives the evaluated solid angle percentage view by the ion, the evaluated temperature, and the contributions to the BBR shift evaluation uncertainty.

Table II. The evaluated solid angle percentage view by the ion, the evaluated temperature, and the contributions to the BBR shift evaluation uncertainty.

| BBR environment | Solid angle percentage | Temperature (K) | Contributions to the total BBR shift uncertainty ($10^{-18}$) |
|---|---|---|---|
| BBR shield (copper) | 33.64% | 82(5) | 0.43 |
| BBR shield (BK7 glass) | 5.45% | 82(5) | 0.07 |
| Ion trap (including the compensation wires) | 14.19% | 82(5) | 0.18 |
| Endcaps | 46.63% | 82(5) | 0.59 |
| Holes on the shield | 0.09% | 293.0(1.5) | 1.46 |
| Total | 100% | | 2.7 |

**Excess micromotion induced shift.** Excess micromotion exists when the ion's position is pushed away from the trap center by the stary field [37], the RF phase difference between the RF electrodes would also give rise to the micromotion [38]. The micromotion would cause both the 2$^{nd}$-order

Doppler shift and the Stark shift. In our experiment, voltages are applied on the compensation electrodes for minimize the micromotion. The micromotion sideband spectroscopy and a CCD camera are used for evaluating the micromotion amplitude. The 2nd-order Doppler shift and the Stark shift due to micromotion would cancel each other by choosing the magic RF trapping frequency [15,18], thus frequency shift uncertainty due to the excess micromotion was kept at a very low level, ~$2\times10^{-19}$.

**Secular motion induced shift.** During the clock run, the trapped ion is firstly laser cooled to a relatively low temperature (lower amplitude of secular motion), then the cooling lasers are blocked for avoiding the light shift, after that the state preparation, the clock interrogation, and the state detection are carried out. The ion' temperature can be evaluated by measuring the amplitude ratio between the clock transition carrier and its 1st-order red secular sidebands [39]. When the cooling lasers are blocked, the ion trap would heat the ion, making the ion's temperature higher with longer state preparation and interrogation time. Both the ion's temperature and the ion trap's heating rate are measured every day. Fig. 4 shows the measured ion's temperature for the cryogenic clock during the clock comparison experiments. The ion's temperature is between 0.65 and 1.17 mK, close to the Doppler cooling limit, we take 0.91(26) mK for the evaluation of the secular motion induced shifts. The secular motion would cause three kinds of shifts: the 2nd-order Doppler shift, the Stark shift, and another different 2nd-order Doppler shift due to the micromotion (secular motion induced). Under the magic RF frequency, the Stark shift, and another different 2nd-order Doppler shift due to the micromotion (secular motion induced) would cancel each other [15,18], taking the evaluated ion temperature, the total secular motion induced shift would be -3.1(9)×$10^{-18}$.

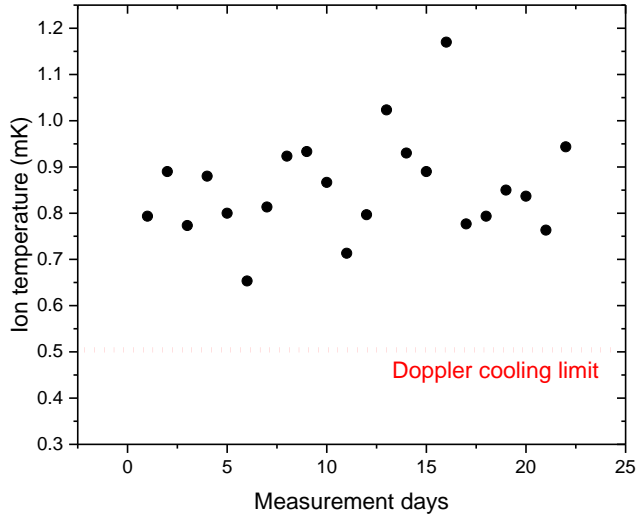

Fig. 4. The measured ion's temperature for the cryogenic clock during the clock comparison, the heating rates are counted in when making the measurements.

**Probe laser induced light shift.** Theoretically, for $^{40}Ca^+$ with a weak probe laser power, the light shift due to the probe laser would be negligible with the hyper-Ramsey methods [14,31]. The ac Stark shift under the Rabi interrogation situation is measured first, by an interleaved comparison measurement increasing the power by 400 times to improve the measurement uncertainty. For a 4-ms-long pulse, the ac Stark shift is measured as 0.2(0.3) Hz, while for an 80-ms-long pulse, the ac Stark shift is measured as 0.53(78) mHz. Taken the measured shift under the Rabi interrogation, the shift for our hyper-Ramsey interrogation is simulated, yielding a light shift of 0.00(4) mHz, or < 1 ×10$^{-19}$ for both the shift and the uncertainty.

**Residual quadrupole shift.** For the $^{40}Ca^+$, the quadrupole shift would be [39]

$$\Delta\nu_{elect-quadr} = -\frac{1}{4}\frac{dE_z}{dz}\Theta(D,J)\frac{[3m_J^2-J(J+1)]}{J(2J-1)}(3cos^2\beta-1) \quad (2)$$

where $dE_z/dz$ is the electric field gradient, $\beta$ is the angle between the magnetic field and the electric field gradient, $\Theta(D,J)$ is the quadrupole moment. Since the electric field gradient is difficult to be precisely controlled or measured, it is difficult to precisely measure the quadrupole moment. Thus, in our experiment, the quadrupole shift is not evaluated by using Eq. (2), while it is eliminated by averaging $m_J$ sub levels [28-30]. However, the quadrupole shift would have a slow drift during the experiment, thus would change a little amount between probing different Zeeman sub levels, causing a residual quadrupole shift. To make the residual quadrupole shift uncertainty even smaller, a specific magnetic field direction is chosen that $3cos^2\beta-1=0$. With this "magic" angle $\beta_0$, the electric quadrupole shift for every Zeeman sub levels would be ~0, and the quadrupole shift drift is about 2 orders of magnitude smaller comparing to our previous work [18,26,30]. The hardware and software for the clock is also upgraded, enabling fast changing the RF applied on the AOMs, it takes only ~0.2 s for switching between different Zeeman sub levels instead of ~8 s. The residual quadrupole shift is greatly suppressed, with an uncertainty of 4×10$^{-19}$.

**1$^{st}$-order Doppler shift.** The photoelectric effect from laser beams may lead to changes in the stray electric field, which results in ion displacement in the ion trap. In the experiment, two laser beams in opposite directions were used for interleaved probing to eliminate such frequency shifts. For every probe pulse, only one beam is frequency tuned to the transition resonance, while the other is tuned to be 1 MHz red detuned. The two beams are well overlapped, and the angle between two opposite propagated beams is evaluated to be < 1 degree. The clock transition frequencies observed in the present experiment using two probe beams differed from each other within 1×10$^{-17}$. Compared to Ref [12], the reason why no obvious 1$^{st}$-order Doppler shift is observed in our experiment is remain unclear, however, one possible explanation is the ultraviolet laser beams may cause a more obvious frequency shift. With our measured ion displacement, the uncertainty in the 1$^{st}$-order Doppler shift was estimated to be 3×10$^{-19}$ with similar method to Ref [12].

**Other systematic shifts** Other systematic shifts are also evaluated including: light shift due to lasers other than the probe laser, uncertainty due to the knowledge of the differential scalar polarizability and the dynamic correction [18], the residual 1$^{st}$-order Zeeman, the 2$^{nd}$-order Zeeman, line pulling, AOM chirp, collisional shift [30, 40], etc. With measurements or calculation, for our cryogenic clock, the shifts are within 1×10$^{-19}$.

**Frequency comparison between the room temperature clock and the cryogenic clock**
**Ion trap used in the room temperature clock.** The ion trap used in the room temperature has a mostly identical design to the NIST's [12,33], the gold coating thickness is made to ~5 μm for achieving smaller trap temperature rise brought by the RF electric trapping field. The thickness of the diamond wafer is ~ 300 μm, while the distance between the RF electrodes and the trapped ion is ~ 400 μm. Very small heating rates is measured for this trap: < 5 quanta/s for our radial secular frequencies of ~2π×3 MHz. Temperature sensors are installed on the vacuum chamber for monitoring the temperature of the vacuum chamber, while no sensors are installed in-vacuum for precisely monitoring the temperature of the ion trap. However, a $CaF_2$ viewport is added for temperature measurements with inferred cameras. When the ion trap is running at an RF power of ~ 2 W, the power used in the clock comparison experiments, the temperature rise of the ion trap is measured as < 0.2 K, limited by the temperature measurement resolution of the camera. Meanwhile, we have built another ion trap, identical to the room temperature clock, but with 3 in-vacuum temperature sensors, for directly measuring the temperature of the ion trap: one is installed at the diamond wafer, the other two are installed at the post for holding the trap. For the sensors on the post, one is ~ 1 cm away from the trap, the other is ~ 10 cm, close to the vacuum chamber. An RF field with the same frequency and amplitude is applied by checking the ion's secular frequencies, a temperature rise of 0.1(1) K is measured. In conclusion, we take 0.2 K as the upper limit of the temperature difference between the ion trap and the vacuum chamber.

**Systematic uncertainty budget for the room temperature clock.** For the room temperature clock, the systematic shift evaluation methods are mostly identical to the cryogenic clock, except for the BBR shift evaluation. The systematic uncertainty budget for the room temperature clock is shown in Table III.

Table III. Systematic uncertainty budget for the room temperature clock. Here only the effects with systematic shifts or uncertainty $>1\times10^{-19}$ is shown.

| Contribution | Fractional systematic shift ($10^{-18}$) | Fractional systematic uncertainty ($10^{-18}$) |
|---|---|---|
| BBR field evaluation (temperature) | 839 | 17 |
| BBR coefficient ($\Delta\alpha_0$) | 0 | 0.3 |
| Excess micromotion | 0 | 0.2 |
| $2^{nd}$-order Doppler (thermal) | -2.5 | 0.7 |
| Residual quadrupole | 0 | 0.4 |
| Servo | 0 | 0.4 |
| $1^{st}$-order Doppler | 0 | 0.3 |
| Total | 836.5 | 18 |

**Clock stability.** During the clock comparison experiments, most of the time the clocks are referenced to the average 3 pairs of transitions with |m|=1/2, 3/2, and 5/2. The Hyper-Ramsey method is used, and the overall clock interrogation time is chosen as 52 ms. We also have tried locking the clocks to the pair of transitions with |m|=1/2 for ~4.5 h, which have the smallest sensitivity to the magnetic field variations, and the overall clock interrogation time is chosen as 92 ms for achieving better stability. Fig. 5 shows the Allan deviation calculated from the clock comparisons. The data have been divided by √2, showing the stability for a single clock [30].

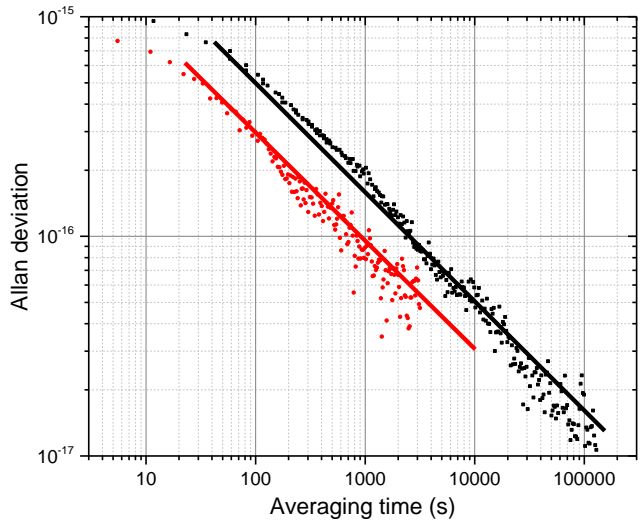

Fig. 5. The Allan deviation calculated from the clock comparisons. The data have been divided by √2, showing the stability for a single clock. The black data points represents when the clocks are referenced to the average 3 pairs of transitions with |m|=1/2, 3/2, and 5/2 for 219 h, the overall clock interrogation time is chosen as 52 ms. The red data points represents when the clocks are referenced to the pair of transitions with |m|=1/2, which have the smallest sensitivity to the magnetic field variations, while the overall clock interrogation time is chosen as 92 ms. The black solid line represents $5\times10^{-15}/\sqrt{\tau}$, while the red solid line represents $3\times10^{-15}/\sqrt{\tau}$.

**Acknowledgments**

We thank Z. Lu, C. Ye, and J. Luo for help and fruitful discussion. This work is supported by the National Key R&D Program of China (Grant Nos. 2017YFA0304401, 2017YFA0304404, 2018YFA0307500, 2017YFF0212003), the Natural Science Foundation of China (Grant Nos. 12022414, 11774388, 11634013), the Strategic Priority Research Program of the Chinese Academy of Sciences (Grant No. XDB21030100), CAS Youth Innovation Promotion Association (Grant Nos. 2018364, Y201963), and the K. C. Wong Education Foundation (Grant No. GJTD-2019-15).

All authors edited and commented on the manuscript. Y. H., B. Z contribute to this work equally. Y. H., B. Z., M. Z., Y. H. H. Z., Z. C., and M.W. designed and performed the experiment, including analyzing the experimental data. B. Z. wrote the program for the experiment. Y. H., H. G., and K. G. supervised the work. Y. H. wrote the manuscript.

**Competing Interests**

The authors declare that they have no competing financial interests. Readers are welcome to comment on the online version of the paper.

**Correspondence**

Correspondence and requests for materials should be addressed to K. G. (email: klgao@apm.ac.cn) or H. G. (email: guanhua@apm.ac.cn)